\documentclass[iop]{emulateapj}
\usepackage{subfigure}
\usepackage{natbib}
\usepackage{url}
\usepackage{appendix}

\begin{document}

\shorttitle{The Evolution of Structure in IRDCs}
\shortauthors{Battersby et al.}

\def\Msun{\hbox{M$_{\odot}$}}
\def\Lsun{\hbox{L$_{\odot}$}}
\def\kms{km~s$^{\rm -1}$}
\def\hcop{HCO$^{+}$}
\def\n2hp{N$_{2}$H$^{+}$}
\def\micron{$\mu$m}
\def\13CO{$^{13}$CO}
\def\etamb{$\eta_{\rm mb}$}	
\def\Inu{I$_{\nu}$}
\def\kapnu{$\kappa _{\nu}$}
\def\ffore{f$_{\rm{fore}}$}
\def\tastar{T$_{A}^{*}$}
\def\nh3{NH$_{3}$}
\def\deg{$^{o}$}
\def\abnh3{$\chi$$_{NH_{3}}$}
\def\H2{H$_{2}$}
\def\arcsec{$^{\prime\prime}$}
\newcommand{\oneone}{(1,1)}
\newcommand{\twotwo}{(2,2)}
\newcommand{\fourfour}{(4,4)}
\input epsf

\title{The Comparison of Physical Properties Derived from Gas and Dust in a Massive Star-Forming Region}

\author{Cara Battersby\altaffilmark{1,2},
John Bally\altaffilmark{1},
Miranda Dunham\altaffilmark{3},
Adam Ginsburg\altaffilmark{1,4},
Steve Longmore\altaffilmark{5},
Jeremy Darling\altaffilmark{1}
}

\altaffiltext{1}{Center for Astrophysics and Space Astronomy, University of Colorado, UCB 389, Boulder, CO 80309}
\altaffiltext{2}{Harvard-Smithsonian Center for Astrophysics, 60 Garden Street, Cambridge, MA 02138, USA}
\altaffiltext{3}{Department of Astronomy, Yale University, New Haven, CT 06520}
\altaffiltext{4}{European Southern Observatory, Karl-Schwarzschild-Strasse 2, D-85748 Garching bei M¬unchen, Germany}
\altaffiltext{5}{Astrophysics Research Institute, Liverpool John Moores University, Twelve Quays House, Egerton Wharf, Birkenhead CH41 1LD, UK}

\begin{abstract}

We explore the relationship between gas and dust in massive star-forming regions by comparing the physical properties derived from each.  We compare the temperatures and column densities in a massive star-forming Infrared Dark Cloud (IRDC, G32.02$+$0.05), which shows a range of evolutionary states, from quiescent to active.
The gas properties were derived using radiative transfer modeling of the (1,1), (2,2), and (4,4) transitions of \nh3 on the Karl G. Jansky Very Large Array (VLA), while the dust temperatures and column densities were calculated using cirrus-subtracted, modified blackbody fits to Herschel data.  We compare the derived column densities to calculate an \nh3 abundance, \abnh3 = 4.6 $\times$ 10$^{-8}$.  In the coldest star-forming region, we find that the measured dust temperatures are lower than the measured gas temperatures (mean and standard deviations T$_{dust, avg}$ $\sim$ 11.6 $\pm$ 0.2 K vs. T$_{gas, avg}$ $\sim$ 15.2 $\pm$ 1.5 K), which may indicate that the gas and dust are not well-coupled in the youngest regions ($\sim$0.5 Myr) or that these observations probe a regime where the dust and/or gas temperature measurements are unreliable.
Finally, we calculate millimeter fluxes based on the temperatures and column densities derived from \nh3 which suggest that millimeter dust continuum observations of massive star-forming regions, such as the Bolocam Galactic Plane Survey or ATLASGAL, can probe hot cores, cold cores, and the dense gas lanes from which they form, and are generally not dominated by the hottest core.

\end{abstract}


\section{Introduction}

Toward massive star and cluster forming regions, we are interested in probing the physical conditions of dense molecular gas clumps, which are highly embedded (A$_{V}$ $\sim$ 10-100).  Hence, we observe at long wavelengths ($\lambda$ $>$ 70 \micron), where the thermal dust emission blackbody spectrum peaks and low energy molecular transitions can be observed.  Understanding the physical conditions, like temperature and column density, at the onset of massive star formation provides crucial constraints for models of star and cluster formation.  

While a variety of molecular gas species (e.g., CO, \nh3, H$_{2}$CO) can be used to trace physical conditions in these dense molecular clumps, \nh3 has the advantage of closely spaced inversion transitions, allowing for observations of multiple transitions in the same observing band, making it a commonly observed species \citep[e.g.,][]{ho83, man92, lon07, pil06, pil11}.  \nh3 has been shown to be a reliable tracer of the mass-averaged gas temperatures to within better than 1 K \citep{juv12}.  The rotational energy states of \nh3 are described by quantum numbers (J,K) and dipole transitions between different K ladders are forbidden.  Therefore, their relative populations  depend only on collisions and are direct probes of the kinetic temperature of the emitting gas.  Each (J,K) rotational energy level is divided into inversion doublets, the (1,1) and (2,2) inversion transitions being most commonly observed \citep[e.g.,][]{rag11, pil06}.  The hyperfine structure of the inversion transitions allows for straightforward measurements of the optical depth of the lines.  The inversion doublet transitions of \nh3 provide a robust tool for measuring gas temperatures and column densities.  

The optically thin thermal emission from dust grains at long wavelengths can also be utilized to derive the physical conditions deep within massive star and cluster forming regions.  In order to derive the temperature and column density of the observed dust, we fit a modified blackbody to the dust emission spectra over a range of wavelengths.  The measured modified blackbody directly traces the thermal emission from dust grains and provides an estimate of the dust temperature and column density, the accuracy of which depends on the number of data points, their uncertainty, and, of course, how well the region can be approximated by the model of a modified blackbody at a single temperature, column density, and dust spectral index.

Gas and dust temperatures and column densities are generally used interchangeably in these dense molecular gas clumps.  In the densest regions of these clumps we expect the gas and dust to be tightly coupled at about n $>$ 10$^{4.5}$ cm$^{-3}$\citep{gol01}.  We note, however, that \citet{you04} found a higher gas-dust energy transfer rate, which may change the derived density threshold of \citet{gol01} by a few 10s of percent.  In this work, we compare the physical properties derived from gas and dust in a massive star-forming Infrared Dark Cloud (IRDC G32.02$+$0.05) that shows a range of evolutionary states \citep{bat14b}.  The gas physical properties are derived using radiative transfer modeling of three inversion transitions of para-\nh3 ((1,1), (2,2), and (4,4)) observed with the Karl G. Jansky Very Large Array (VLA) by \citet{bat14b}.  The dust physical properties are derived using cirrus-subtracted modified blackbody fits data from the Herschel Infrared Galactic Plane Survey \citep[Hi-GAL][]{mol10} using the method described in \citet{bat11}.  The column densities derived from each tracer are compared with column densities derived using 1.1 mm dust emission data from the Bolocam Galactic Plane Survey \citep[BGPS,][]{gin13, agu11, ros10}, 8 \micron~dust absorption data from the Galactic Legacy Mid-Plane Survey Exraordinaire \citep[GLIMPSE,][]{ben03} using the method from \citet{bat10}, and $^{13}$CO emission data from the Boston University-Five College Radio Astronomy Observatory Galactic Ring Survey \citep[BU-FCRAO GRS or just GRS][]{jac06}.

In \S \ref{sec:data} we summarize the data and methods used to derive physical properties.  In \S \ref{sec:abundance} we calculate the abundance of \nh3 in this IRDC and compare the column densities derived from each tracer of the molecular gas.  \S \ref{sec:comp} presents a comparison of the properties derived from gas with those derived from dust.  We compare the high-resolution \nh3 observations with lower-resolution observation from the Green Bank Telescope (GBT) in \S \ref{sec:gbt}.  Finally, in \S \ref{sec:bgps}, we forward model the high-resolution gas temperatures and column densities derived from the VLA to derive the millimeter fluxes that would be observed with the BGPS, allowing us to explore the high-resolution ($\sim$ 0.1 pc) nature of pc-scale dense, molecular clumps.  We conclude in \S \ref{sec:conc}.
  
\section{Data}
\label{sec:data}

%made in ds9
\begin{figure*}
\centering
\subfigure{
\hspace{-0.75in}
\includegraphics[width=0.49\textwidth]{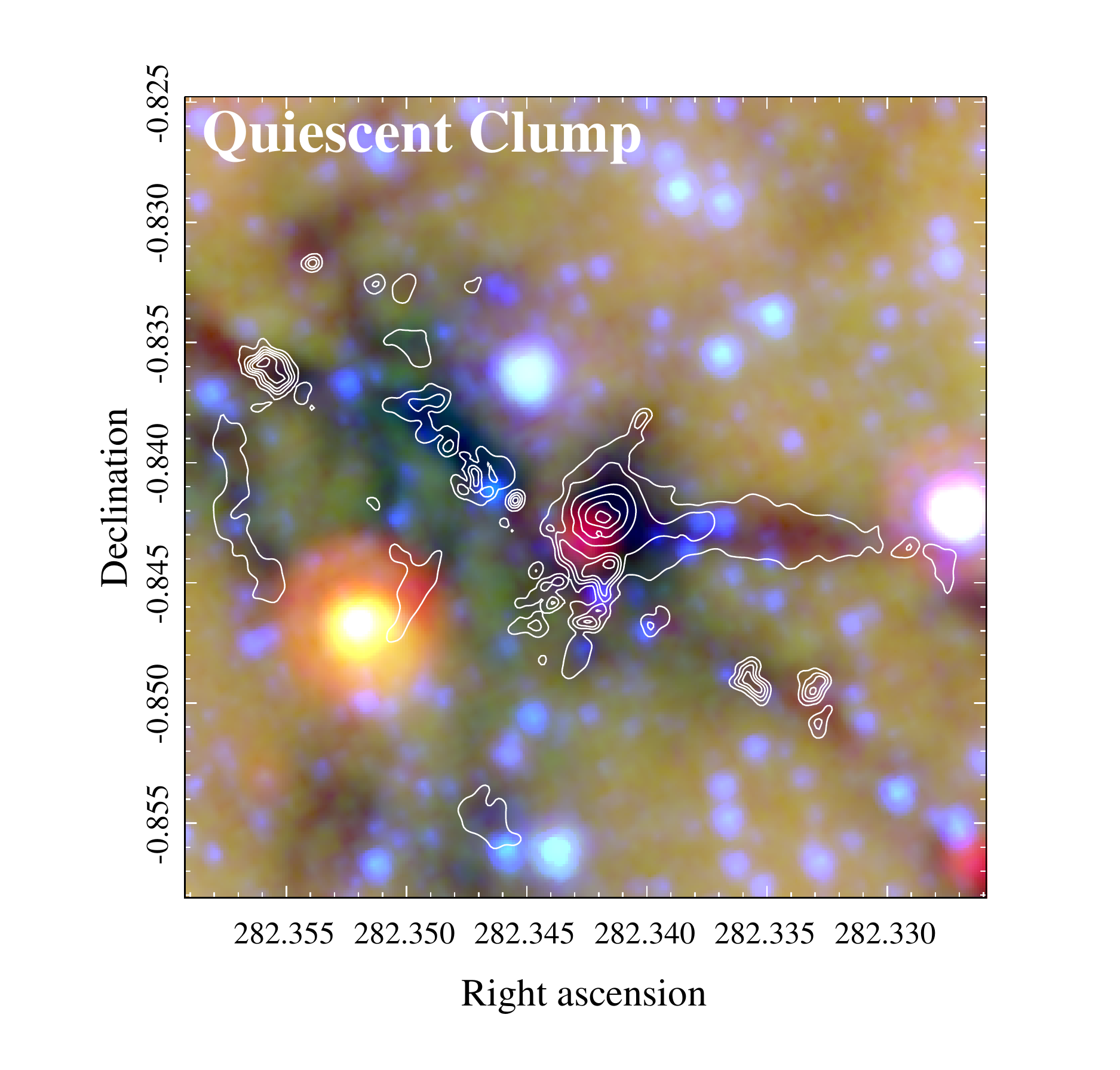}}
\hspace{-0.5in}
\subfigure{
\includegraphics[width=0.49\textwidth]{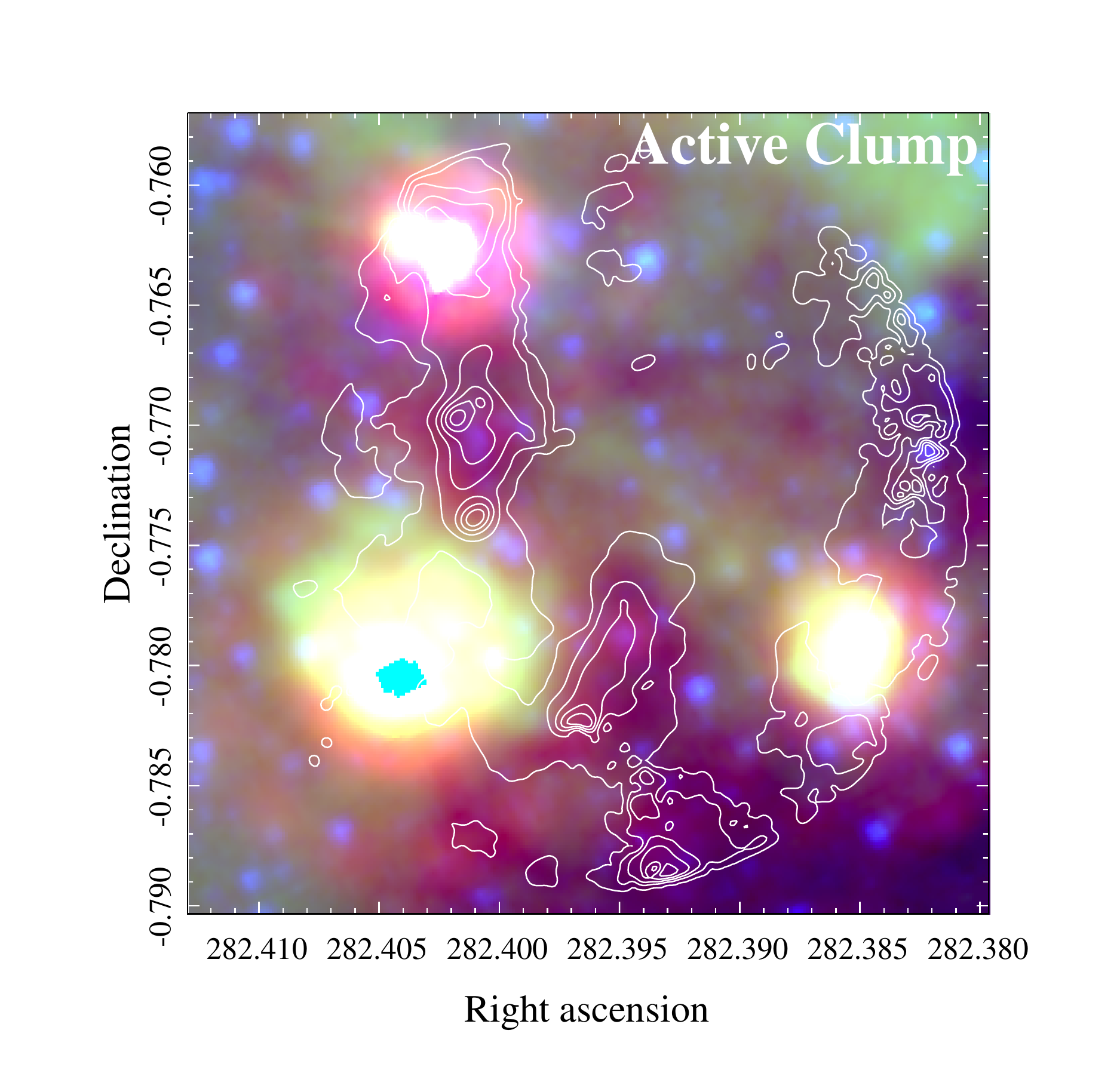}}
 \\
\caption{Images depicting the two clumps studied in this work.  The images are 3-color in the mid-IR (MIPSGAL, GLIMPSE, Red: 24 \micron, Green: 8 \micron, Blue: 4.5 \micron) with column density contours derived from \nh3 (six linearly spaced contours, in the quiescent clump from 8$\times$10$^{21}$ cm$^{-2}$ to 10$^{23}$ cm$^{-2}$ and in the active clump, from 10$^{22}$ cm$^{-2}$ to 3$\times$10$^{23}$ cm$^{-2}$).}
\label{fig:glmimage}
\end{figure*}

\subsection{VLA \nh3}
The observations and radiative transfer modeling used to derive \nh3 gas temperatures and column densities are presented and explained in detail in \citet{bat14b}.

The (1,1), (2,2), and (4,4) inversion transitions of para-\nh3 were observed toward two clumps within IRDC G32.02$+$0.05 with the National Radio Astronomy Observatory\footnote[1]{The National Radio Astronomy Observatory is a facility of the National Science Foundation operated under cooperative agreement by Associated Universities, Inc.} Karl G. Jansky Very Large Array (VLA).  We observed two clump locations within the IRDC G32.02$+$0.06, an active clump 
([$\ell$, b] = [32.032\deg, $+$0.059\deg]) and a quiescent clump ([$\ell$, b] = [31.947\deg, $+$0.076\deg]), see Figure \ref{fig:glmimage}. 
The active clump displays signs of active star formation including a 6.7 GHz methanol maser \citep{pes05}, 8 and 24 \micron~emission as well as radio continuum emission \citep[see][]{bat14b, whi05, hel06} indicative of Ultra-Compact HII Regions.  The quiescent clump doesn't show any of those star formation signatures, except for possible association with a faint 24 \micron~point source.  The final beam FWHM produced by the model was about 4.4\arcsec~\citep[$\sim$0.1 pc at the adopted distance of 5.5 kpc,][]{bat14b} and an RMS noise of about 6 mJy/beam.

The gas physical properties were derived from the inversion transitions using radiative transfer modeling of the lines.  The ammonia lines were fit with a Gaussian line profile to each hyperfine component simultaneously with frequency offsets fixed.  The fitting was performed in a Python routine translated from Erik Rosolowsky's IDL fitting routines  \citep[Section 3 of ][]{ros08}.  The model was used within the framework of the {\tt pyspeckit} spectral analysis code package \citep[][\url{http://pyspeckit.bitbucket.org}]{gin11}.  Typical statistical errors are in the range $\sigma$(T$_{K}$) $\sim$ 1-3 K for the kinetic temperature, and $\sigma\left(\log[N(NH_{3})]\right)\lesssim0.05$ (about 10\% in N(\nh3)) for the column density of ammonia.  See \citet{bat14b} for more details.

\subsection{Dust Continuum Column Density}
\label{sec:dust_col}
\subsubsection{Herschel Infrared Galactic Plane Survey}
The Herschel Infrared Galactic Plane Survey, Hi-GAL \citep{mol10}, is an Open Time Key Project of the Herschel Space Observatory \citep{pil10}. Hi-GAL has performed a 5-band photometric survey of the Galactic Plane in a $|b| \leq1^{o}$ -wide strip from -70$^{o}$ $\leq$ $\ell$ $\leq$ 70$^{o}$ at 70, 160, 250, 350, and 500 \micron~using the PACS \citep[Photodetector Array Camera and Spectrometer,][]{pog10} and SPIRE \citep[Spectral and Photometric Imaging Receiver,][]{gri10} imaging cameras in parallel mode.  Data reduction was carried out using the \textit{Herschel} Interactive
Processing Environment \citep[HIPE,][]{ott10} with custom reduction scripts that deviated considerably from the standard processing for PACS \citep{pog10}, and to a lesser extent for SPIRE \citep{gri10}.  A more detailed description of the entire data reduction procedure can be found in \citet{tra11}.  A weighted post-processing on the maps \citep{pia13} has been applied to help with image artifact removal.

We use the Hi-GAL data to derive dust continuum column densities and temperatures at 36\arcsec~resolution using pixel-by-pixel modified blackbody fits to the SPIRE 250, 350, and 500 \micron~data.  Generally, the PACS 160 \micron~data are also part of the modified blackbody fit \citep[as in][]{bat11}, however, the quiescent clump has an exceptionally cold temperature and relatively low column density, such that the PACS 160 \micron~point is lower than the background (this happens in less than $\sim$5\% of pixels) and so that point is not included.  The fits are very similar both qualitatively and quantitatively with and without the PACS 160 \micron~point within about 10\% in column density and about 3\% in temperature).  

The modified blackbody fits are performed on data that has had the cirrus foreground subtracted using an iterative routine discussed in detail in \citet{bat11}.  The modified blackbody fits assume a spectral index, $\beta$, of 1.75, a gas to dust ratio of 100, a mean molecular weight of 2.8 \citep[e.g.,][]{kau08}, and the \citet{oss94} MRN distribution model with thin ice mantles that have coagulated at 10$^{6}$ cm$^{-3}$ for 10$^{5}$ years for the dust opacity.  As in \citet{bat11}, we fit a power-law to the \citet{oss94} dust opacity to have a continuous dust opacity as a function of wavelength.  This power-law fit gives us a dust opacity of 5.4 cm$^{2}$/g at 500 \micron~\citep[compared with 5.0 cm$^{2}$/g from][]{oss94} and a value of $\beta$ of 1.75.  %The cirrus f

\subsubsection{Bolocam Galactic Plane Survey}
We utilize the 1.1 mm dust continuum emission from version 2 of the Bolocam Galactic Plane Survey \citep[BGPS,][]{gin13, agu11, ros10} to estimate the isothermal column densities.  We derive column density maps at 33\arcsec~resolution assuming the dust temperature from the corresponding pixel in Hi-GAL.  The column density is given by
\begin{equation}
N(H_{2}) = \frac{S_{\nu} }
{\Omega_{B}\kappa_{\nu}B_{\nu}(T) \mu_{H_{2}}m_{H}}
\end{equation}
\begin{equation}
\label{eq:dust_col}
N(H_{2})_{1.1 mm}= 2.20\times10^{22}(e^{13.0/T}-1)S_{\nu}~\rm cm^{-2}
\end{equation}
where S$_{\nu}$ is the source flux (in Jy in Eq. \ref{eq:dust_col}), B$_{\nu}$(T) is the Planck function at dust temperature T, $\Omega_{B}$ is the beam size, and $\mu_{H_{2}}$ is the mean molecular weight for which we adopt a value of $\mu_{H_{2}}$ = 2.8 \citep{kau08}.  We adopt a value of $\kappa_{1.1 mm}$ = 1.14 cm$^{2}$ g$^{-1}$, interpolated from the tabulated values of dust opacity from the \citet{oss94} MRN distribution model with thin ice mantles that have coagulated at 10$^{6}$ cm$^{-3}$ for 10$^{5}$ years.  This same model for dust opacity was used for the Hi-GAL and 8 \micron~extinction derived column densities.  We assume a gas-to-dust ratio of 100.  We use the dust temperatures derived from Hi-GAL in each pixel above in the calculation of the column densities.  
The same analysis is possible with ATLASGAL 870\micron \citep{sch09} or SCUBA-2 850um data \citep{cha13}, which like the Bolocam data trace column density and are relatively insensitive to the dust temperature.  However, at the time the analysis was performed, no data sets from these instruments was available.

\subsection{Extinction Derived Column Density}
\label{sec:ext_col}
Dense clumps of dust and molecular gas will absorb the bright mid-IR Galactic background and appear as dark extinction features in the mid-IR \citep[e.g.,][]{car98, ega98, per09}.  We use data from the Galactic Mid-Plane Survey Extraordinaire \citep[GLIMPSE;][]{ben03} to derive column density maps from the extinction of the dense clumps at 8 \micron~at 2\arcsec~resolution.
The extinction mass and column density are calculated using the \citet{but09} method with the correction as applied in \citet{bat10}.  We use a dust opacity ($\kappa_{8 \mu \rm m}$ = 11.7 cm$^{2}$ g$^{-1}$) from the same \citet{oss94} model as above,
which is a reasonable model for the cold, dense environment of an IRDC.  This and the assumption of a gas-to-dust ratio of 100 is consistent with the opacity used for our dust emission column density estimates.  This method uses an 8 \micron~emission model of the Galaxy and the cloud distance to solve the radiative transfer equation for cloud optical depth.  The expression for the gas surface density is
\begin{equation}
\Sigma = -\frac{1}{\kappa_{\nu}} \rm ln \Bigg[\frac{(\rm s +1)I_{\nu1,\rm obs} - (\rm s + \rm f_{\rm fore})I_{\nu0,\rm obs}}{(1-\rm f_{\rm fore})I_{\nu0,\rm obs}}\Bigg]
\end{equation}
where $\kappa$$_{\nu}$ is the dust opacity, I$_{\nu1,\rm obs}$ and I$_{\nu0,\rm obs}$ are the observed intensities in front of and behind the cloud respectively, s is the IRAC scattering correction \citep[0.3, see][]{bat10}, and f$_{\rm fore}$ is the fraction of total emission along the line of sight produced in the foreground of the cloud (0.3 for this cloud at the kinematic distance of 5.5 kpc).
  We compare this estimate with the method using the 8 \micron~extinction derived column density from \citet{per09}.  The \citet{per09} method assumes that most of the observed 8 \micron~emission is local to the cloud, so a distance is not required.  For this particular source, the column densities derived using the \citet{per09} method are about 15\% higher than the \citet{but09} method, with very little scatter.  We use the \citet{but09} method for the remainder of the discussion.

\subsection{GRS \13CO Column Density}
\label{sec:co_col}
We utilize data taken as part of the Boston University Five College Radio Astronomy Observatory \citep[BU-FCRAO GRS or just GRS,][]{jac06} of the $^{13}$CO J=1-0 transition at 46\arcsec~resolution to calculate a column density assuming a gas temperature from the corresponding pixel in the \nh3 gas temperature map.
In the optically thin, thermalized limit, the H$_{2}$ column density derived from \13CO is given by
\begin{equation}
N(H_{2}) = \frac{8\pi k\nu^{3}X_{^{13}CO}}{3c^{3}hB_{J}A_{10}}
(1-e^{-h\nu/kT_{ex}})^{-1}\int \! T_{mb} \, dv
\end{equation}
where $\nu$ is the frequency of the \13CO J=1-0 transition, A$_{10}$ is the Einstein A coefficient of \13CO from state J=1 to J=0, T$_{ex}$ is the excitation temperature, B$_{J}$ is the rotation constant, and X$_{^{13}CO}$ is the abundance fraction of \13CO to \H2.  We adopt a value of $^{12}$CO / \13CO of 58 from \citet{luc98}, and a value of $^{12}$CO / \H2 of 10$^{-4}$, a value of 55.101038 GHz for B$_{J}$, and standard NIST values for all constants and spectral transition values.  This expression then reduces to 
\begin{equation}
N(H_{2}) = 1.45 \times 10^{20} \frac{\int \! T_{mb} \, dv}{1 -
  e^{-5.29/T}}\rm~cm^{-2}, 
\end{equation}
assuming temperatures in K and velocity in \kms.
For more details on how we derive a column density estimate from $^{13}$CO see \S 3.5 of \citet{bat10}.

%%%%%%%%%%%%%%
%% Abundance Figure %%
%%%%%%%%%%%%%%
% Made using 'plot_fit_abundance_temp.pro'
% and now made using 'fit_abundance_temp.pro'
% old ones:
% 'coldclump_col_comp_temp-eps-converted-to.pdf'
% 'hotclump_col_comp_temp-eps-converted-to.pdf'
\begin{figure*}
  \centering
  \subfigure{
    \includegraphics[width=0.49\textwidth]{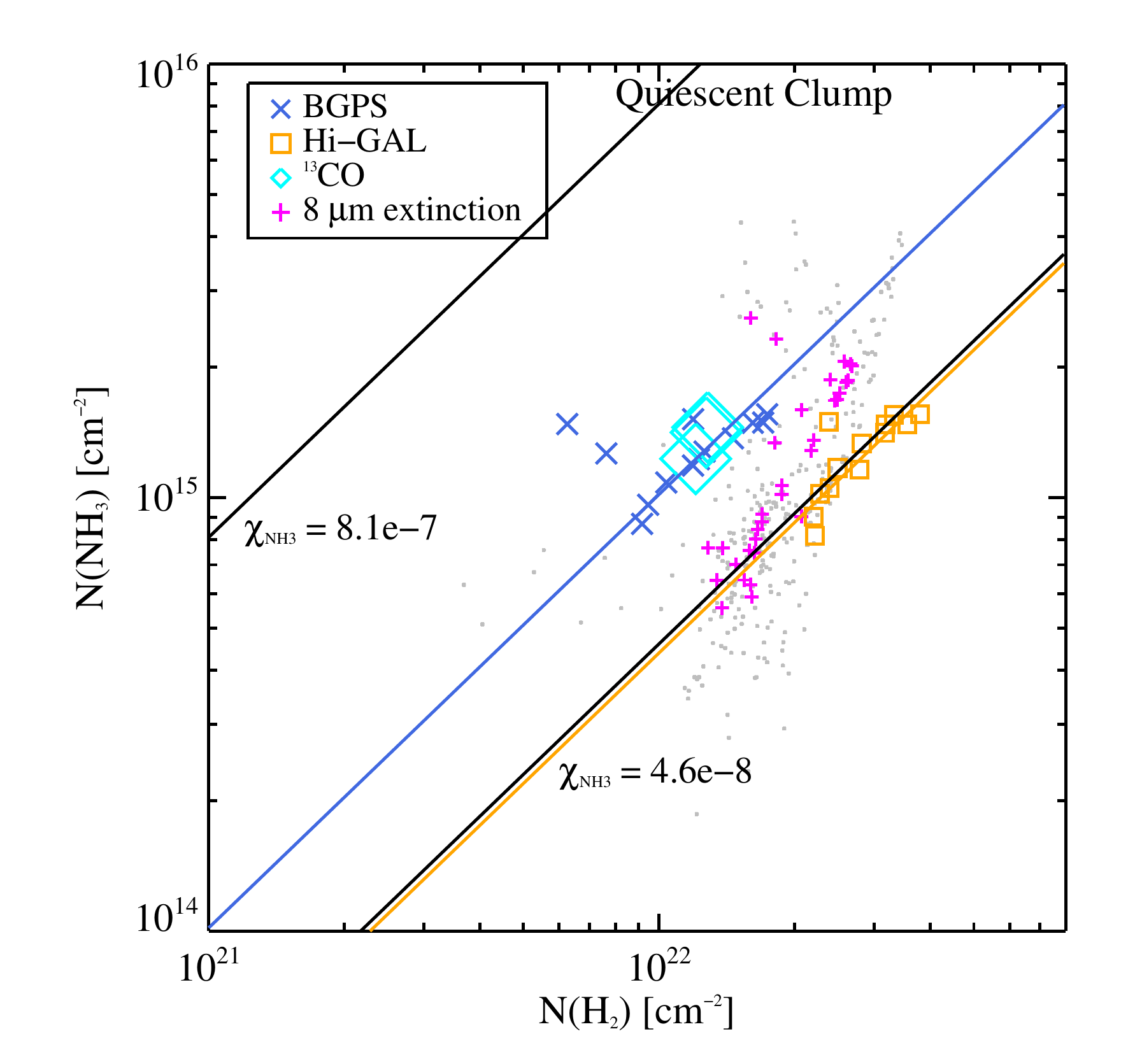}}
  \hspace{-0.25in}
  \subfigure{ 
    \includegraphics[width=0.49\textwidth]{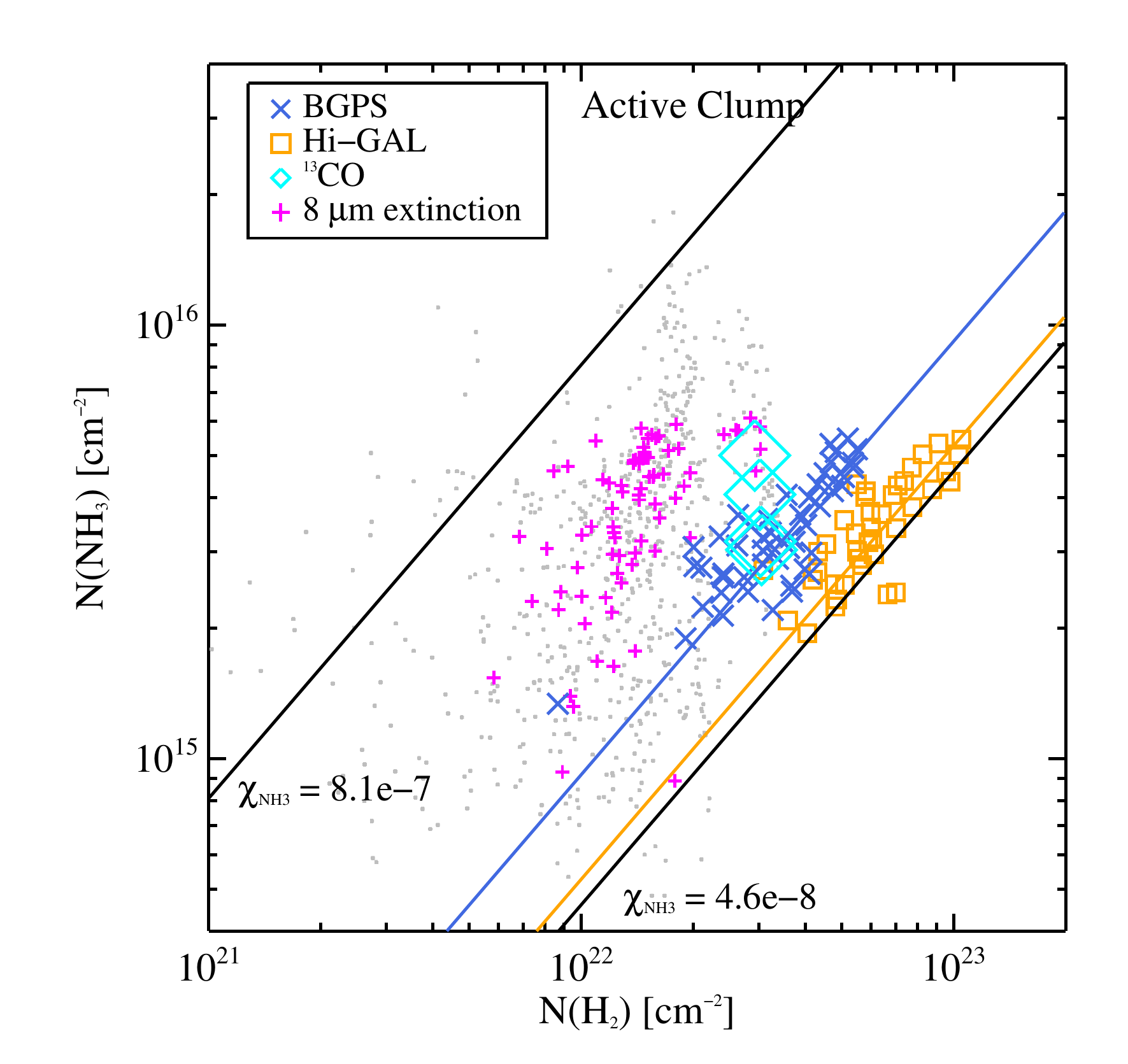}}
  \caption{The correlation of column densities derived from various independent data-sets vs. the ammonia column density for the abundance calculation.  Column densities are derived from Hi-GAL and BGPS 1.1 mm dust continuum emission (using dust temperatures derived from Hi-GAL), \13CO integrated intensity, and 8 \micron~extinction are plotted on the x-axis against the \nh3 integrated intensity column density on the y-axis.  The \textit{left} plot is pixels within the quiescent clump, while the \textit{right} is the active clump.  Both dust continuum emission tracers show good correlation with the \nh3 emission, however, with a systematic offset.  The 8 \micron~extinction estimate shows excellent morphological agreement on large scales but the correlation on small scales is highly scattered (gray points show all data, magenta is their average over 15 pixels for comparison).  The details of these plots are discussed in \S \ref{sec:abundance}.  Linear fits are shown for the BGPS (in blue) and Hi-GAL (in orange) points.   We also show two abundance lines for comparison, $\chi_{NH_{3}}$ = 8.1 $\times$ 10$^{-7}$ from \citet{rag11} and  $\chi_{NH_{3}}$ = 4.6 $\times$ 10$^{-8}$ from this work and \citet{pil06}. 
  }
  \label{fig:abundance}
\end{figure*}

\section{Abundance and the Correlation of N(\nh3) with N(H$_{2}$)}
\label{sec:abundance}
An abundance measurement of \nh3 (\abnh3 = $\frac{N(NH_{3})}{N(H_{2})}$) can be derived from a comparison of the derived \nh3 column densities to the \H2 column densities from a variety of independent measurements: dust continuum emission, dust extinction, and \13CO line emission.  The dust continuum column densities are derived as described in \S \ref{sec:dust_col} using Hi-GAL and BGPS (36\arcsec~and 33\arcsec~resolution, respectively) with dust temperatures from Hi-GAL \citep[]{bat11}.  The dust extinction derived column density is calculated using 8 \micron~extinction as described in \S \ref{sec:ext_col} with 2\arcsec~resolution.  We also utilize N(\H2) derived from \13CO (as described in \S \ref{sec:co_col}) using the \nh3 gas temperatures, despite the \13CO's much lower resolution at 46\arcsec. 

For a pixel-by-pixel comparison with the \nh3 column density maps in the quiescent and active clump, we convolve and re-grid the higher resolution data (the \nh3 column density maps in all cases except the 8 \micron~extinction) to match the lower resolution data for each clump.  The pixel-by-pixel comparison is shown in Figure \ref{fig:abundance}.  While there is a good deal of scatter, there also appear to be systematic differences in the column densities derived from the different methods.  We comment briefly on the magnitude and possible reasons for these differences below.  These systematic disagreements should help to give a flavor for the uncertainties involved in such measurements; \citet{bat10} conclude that typical mass uncertainties using such tracers are about a factor of two. %80\% - 200\% 

For each clump and each H$_{2}$ tracer we fit a line between N(\nh3) and N(H$_{2}$).  The data were fit using the total least squares method, which takes into account errors on both the x and y data points (unlike ordinary least squares, which assumes no error on the x data).  The errors were assumed to be 50\% of the column values for both the \nh3 and total column density.  We first fit a line with the \nh3 y-intercept fixed at zero (the typical method for determining abundances), then fit a line allowing for a non-zero y-intercept.  We only include points with N(\H2) between 10$^{20}$ and 10$^{24}$ cm$^{-2}$ which also fit the ``mask" criteria applied to all the \nh3 data as described in the companion paper, \citet{bat14b}.  We exclude a handful of noisy pixels near the edges of the maps.  The results of the fits are shown in Table \ref{table-abd}.   

\textit{Hi-GAL: } The Hi-GAL column densities show a good correlation with the \nh3 data.  The average derived \nh3 abundance is about 4.6 $\times$ 10$^{-8}$, or with a non-zero y-intercept, about 3.3 $\times$ 10$^{-8}$ with an \nh3 intercept of about 3-9 $\times$ 10$^{14}$ cm$^{-2}$.  Rather than harboring any physical meaning, this non-zero intercept likely indicates that the abundance relationship is not well described by a single linear fit for all column densities. 

\textit{BGPS: } 
%re-done with v2
The BGPS column densities also show a good correlation with the \nh3 column densities.  The average abundance is about 8.6 $\times$ 10$^{-8}$ fixing the y-intercept at zero, or 5.7 $\times$ 10$^{-8}$ with an intercept of about 4 $\times$ 10$^{14}$ cm$^{-2}$.  There is a systematic offset of the BGPS column densities from the Hi-GAL measured column densities of about 2, which can be plausibly explained (see \S \ref{sec:bgpshigal}).

\textit{8 \micron~Dust Extinction: } The \nh3 and 8 \micron~dust extinction show excellent morphological agreement \citep[as seen in the companion paper,][]{bat14b}; the masks created using a column density threshold for each of them are nearly identical.  However, comparing their column densities within those masks reveals a great deal of scatter.  All the data points are plotted in Figure \ref{fig:abundance} in gray, and their averages (on 15\arcsec~scales to show less scatter) in magenta.  A few interesting things to note: 
1) we do not find any extinction derived column densities greater N(H$_{2}$) $\sim$ 3.5 $\times$ 10$^{22}$ cm$^{-2}$ with a somewhat significant cutoff at this column density.  This cutoff is likely due to the increasing optical depth of the 8 \micron~emission which would saturate the extinction measurement.  Using our dust opacity at 8 \micron~of 11.7 gcm$^{-2}$ (see \S \ref{sec:ext_col}), $\tau$$_{8 \mu m}$ = 1 at N(H$_{2}$) $\sim$ 2 $\times$ 10$^{22}$ cm$^{-2}$, which, given reasonable systematic error estimates, can easily can explain the observed cutoff.
2) the active clump is a bit warmer and has a brighter local 8 \micron~background which fills in some of the absorption, explaining the low extinction derived column densities in that region, and 3) variations in the \nh3 abundance, 8 \micron~background, and/or dust opacity are likely responsible for the large scatter in the extinction derived column densities, despite their high resolution.  The ratio of extinction to BGPS column densities is about the same as found in \citet{bat10}, with a ratio of 0.5 (was 0.7, now 0.5 with the 1.5 correction factor to BGPS v2) in the active clump and a ratio of about 1.3 (was 2.0, now 1.3 with 1.5 correction factor to BGPS v2) in the quiescent clump.

\textit{GRS: } While we wouldn't expect to find a very strong correlation in column density over the few resolution elements the GRS $^{13}$CO has over the VLA field of view, the comparison of the overall column densities is informative.  The total column density derived from $^{13}$CO GRS is about half that from Hi-GAL and slightly less than the BGPS (about 90\%).  Whether this indicates CO depletion, optically thick $^{13}$CO, or some other physical effect cannot be determined from these data alone.  This is representative of typical systematic errors of a factor of two in column determinations.

We adopt the average Hi-GAL derived abundance of \abnh3 = 4.6 $\times$ 10$^{-8}$ throughout the remainder of this analysis.  Since the Hi-GAL column densities are derived simultaneously with the dust temperature, and utilize data at multiple wavelengths to perform full modified blackbody fits, we consider that these column densities may be more robust than those determined at a single wavelength with an assumed temperature.  We note, however, that BGPS-derived column densities would be less susceptible to gradients in the dust temperature.
Additionally, the conditions probed with the Hi-GAL data (dust continuum emission from cold, dense clumps) are roughly the same conditions probed with \nh3 (dense gas emission from cold, dense clumps), making the Hi-GAL data a reasonable choice for the derivation of \nh3 abundance.  We do not see significant evidence of \nh3 depletion.  If the difference in the measured Hi-GAL \nh3 abundance between the quiescent and active clump is due to depletion, then the \nh3 is depleted by about 30\% in column density in the quiescent clump, see Figure \ref{fig:depletion}.

Our measured average abundance, \abnh3 = 4.6 $\times$ 10$^{-8}$, is very close to that derived by \citet{pil06} in IRDCs. 
Additionally, \citet{dun11b} studied the abundance of ammonia as a function of Galactocentric radius and found a decrease in abundance of a factor of 7 from a radius of 2 to 10 kpc.  Given the Galactocentric radius of the IRDC studied here (4.7 kpc), the relationship determined by \citet{dun11b} predicts an abundance of $4.5\times10^{-8}$, which is in agreement with both the abundance determined in this study and the recent abundance determination toward this cloud by \citet{chi13}.  One major difference is that \citet{pil06, dun11b, chi13} use single-dish \nh3 observations of two \nh3 lines, while we use interferometric observations of \nh3 modeled with three lines, all para-\nh3.  Our higher resolution observations are probing smaller spatial scales and therefore deriving higher average column densities towards the densest regions, however, we are also insensitive to spatial scales above 66\arcsec.  
The high resolution observations are probing the densest gas on small spatial scales, while the single dish observations are probing all the gas in the beam.  The fact that the abundance measurements are consistent between the observations suggests that either this abundance ratio is constant for the range of diffuse to dense gas that is being probed by both or that the \nh3 and dust we are probing with both measurements arises primarily from dense gas on small scales.
These abundances are also consistent with previous literature values of 3 $\times$ 10$^{-8}$ and 6 $\times$ 10$^{-8}$ from \citet{wan08} and 3 $\times$ 10$^{-8}$ from \citet{har93}.
%largest angular scale from: https://science.nrao.edu/facilities/evla/docs/manuals/oss-2013a/performance-of-the-evla/resolution
Our adopted abundance, \abnh3 = 4.6 $\times$ 10$^{-8}$ differs from the interferometrically derived 8 \micron~extinction based abundance of 8.1 $\times$ 10$^{-7}$ by \citet{rag11} in IRDCs, but given the large scatter in our extinction-derived abundances (average of 1.1 $\times$ 10$^{-7}$), the disagreement is not too surprising.

\subsection{The BGPS and Hi-GAL Discrepancy}
\label{sec:bgpshigal}

In comparing the column densities derived from various datasets, we find a systematic offset of about 2 from Hi-GAL to BGPS derived column densities (Hi-GAL derived column densities are, on average, about 2 $\times$ higher).  We suggest that a handful of smaller effects can explain this offset.
Because the atmospheric subtraction used for BGPS filters out large scale structure, Hi-GAL is more sensitive to larger spatial scales.  Due to this difference in derivation of large scale, despite the background subtraction, we might still expect slightly lower values in the BGPS due to spatial filtering (perhaps of order 20\% ).  We have checked and confirmed that the difference does not arise from the convolutions, or re-gridding of the data.  While both use the same model for the dust opacity \citep{oss94}, the power-law fit opacity used for Hi-GAL extrapolated from \citet{oss94} is slightly higher (0.0135 using a power-law fit and $\beta$=1.75 vs. 0.0114, linearly interpolating tabulated values near 1.1 mm, about 1.2 $\times$ higher) at 1.1 mm than the tabulated value used in the BGPS column density.  The Hi-GAL procedure required a dust opacity that was a continuous function of frequency so we used a power-law fit to \citet{oss94} dust opacity rather than the tabulated value.  Additionally, the spectral index $\beta$ may be steeper than the assumed value of 1.75.  If we assume that about 20\% of the total flux is lost in BGPS due to spatial filtering and use the Hi-GAL power-law extrapolated opacity at 1.1 mm, and assume a spectral index of $\beta$ slightly steeper than 2 (about 2.25) then we can bring the two column densities into agreement.  The large systematic offset can be plausibly explained by a few smaller effects that all push the BGPS column density to lower values.  

%%%%%%%%%%%%%%%%%%%%%%%%%%%%%%%%%%%%%%%%%%%%%%%%%%%%%%%%%%%%%%%%%%%%%%%%%%%%%%%%%%%%%%%%%%%%%%%
% ABUNDANCE MEASUREMENTS
%%%%%%%%%%%%%%%%%%%%%%%%%%%%%%%%%%%%%%%%%%%%%%%%%%%%%%%%%%%%%%%%%%%%%%%%%%%%%%%%%%%%%%%%%%%%%%%

\begin{deluxetable*}{llccccc}
\tabletypesize{\scriptsize}
\tablecaption{Abundance Measurements \label{table-abd}}
\tablewidth{0pt}
%\rotate
\tablehead{
\colhead{} &
\colhead{Comparison of N(NH$_{3}$)} & 
\colhead{\abnh3\tablenotemark{a, b}} &
\colhead{\abnh3 with intercept\tablenotemark{c,b}} &
\colhead{\nh3 intercept\tablenotemark{b,d}} \\
\colhead{} &
\colhead{with N(\H2) from} &
\colhead{} & \colhead{} & \colhead{cm$^{-2}$} }
\startdata

%\multicolumn{2}{l}{Quiescent Clump}  &  &  &  \\
%\cline{1-2}
%  &  Hi-GAL                          &  4.3e-8 $\pm$ 0.2e-8  &  4e-8 $\pm$ 1e-8  &  -0.4e14 $\pm$ 3.6e14 \\
%  &   BGPS                           &  8.7e-8 $\pm$ 0.5e-8  &  5e-8 $\pm$ 2e-8  &  6e14 $\pm$ 3e14 \\
%  &  8 \micron~extinction  &  5.3e-8 $\pm$ 0.2e-8  &  10e-8 $\pm$ 1e-8  &  -7e14 $\pm$ 2e14 \\
%  &  GRS \13CO                 &  10e-8 $\pm$ 1e-8  &  3e-7 $\pm$ 4e-7  &  -2e15  $\pm$ 5e15 \\

%\multicolumn{2}{l}{Active Clump}  &  &  &  \\
%\cline{1-2}
%  &  Hi-GAL                         &  4.8e-8 $\pm$ 0.2e-8  &  4.1e-8 $\pm$ 0.6e-8  &  4e14 $\pm$ 3e14 \\
%  &   BGPS                          &  7.9e-8 $\pm$ 0.2e-8  &  6.8e-8 $\pm$ 0.7e-8  &  4e14 $\pm$ 2e14 \\
%  &  8 \micron~extinction  &  1.67e-7 $\pm$ 0.04e-7  &  10e-8 $\pm$ 1e-8  &  10e14 $\pm$ 2e14 \\
%  &  GRS \13CO                 &  1.1e-7 $\pm$ 0.1e-7  &  -1.6e-6 $\pm$ 0.9e-6  &  5e16 $\pm$ 3e16 \\
%\enddata

% from new abundance fits using MCMC method -- June 24, adam.ginsburg@colorado.edu

\multicolumn{2}{l}{Quiescent Clump}  &  &  &  \\
\cline{1-2}
  &  Hi-GAL                          &  4.0e-8 $\pm$ 0.2e-8  & 3.0e-8 $\pm$ 0.7e-8 & 3e14 $\pm$ 2e14 \\
  &   BGPS                           & 8.6e-8  $\pm$ 0.8e-8  & 3.9e-8 $\pm$ 0.7e-8 & 4.8e14 $\pm$ 0.7e14 \\
  &  8 \micron~extinction  &  4.64e-8 $\pm$ 0.09e-8  & 9.3e-8 $\pm$ 0.4e-8 & -7.4e14 $\pm$ 0.7e14 \\
  &  GRS \13CO                 & 9.5e-8  $\pm$ 0.2e-8  & 1.9e-7 $\pm$ 0.6e-8 & -1.2e15 $\pm$ 0.8e15 \\

\multicolumn{2}{l}{Active Clump}  &  &  &  \\
\cline{1-2}
  &  Hi-GAL                          & 5.3e-8  $\pm$ 0.2e-8  & 3.6e-8 $\pm$ 0.6e-8 & 9e14 $\pm$ 3e14 \\
  &   BGPS                           & 8.7e-8  $\pm$ 0.3e-8  & 7.5e-8 $\pm$ 0.8e-8 & 4e14 $\pm$ 2e14 \\
  &  8 \micron~extinction  & 2.19e-7  $\pm$ 0.04e-7  & 2.46e-7 $\pm$ 0.08e-7 & -2.3e14 $\pm$ 0.7e14 \\
  &  GRS \13CO                 &  1.1e-7 $\pm$ 0.2e-7  & -1.8e-6 $\pm$ 0.6e-6 & 6e16 $\pm$ 2e16 \\

\enddata

\tablenotetext{a}{NH$_{3}$ Abundance: linear fit
  to data without a y-intercept}
\tablenotetext{b}{Assuming a column density error of 20\%}
\tablenotetext{c}{NH$_{3}$ Abundance: linear fit
  to data with a y-intercept}
\tablenotetext{d}{NH$_{3}$ Column density intercept of the fit to the abundance}
\end{deluxetable*}

%%%%%%%%%%%%%%%%%%
%% Hi-GAL NH3 Comp Figures  %%
%%%%%%%%%%%%%%%%%%
%made using 'higal_nh3comp.pro'
\begin{figure*}
\centering	
\subfigure{
\includegraphics[width=0.49\textwidth]{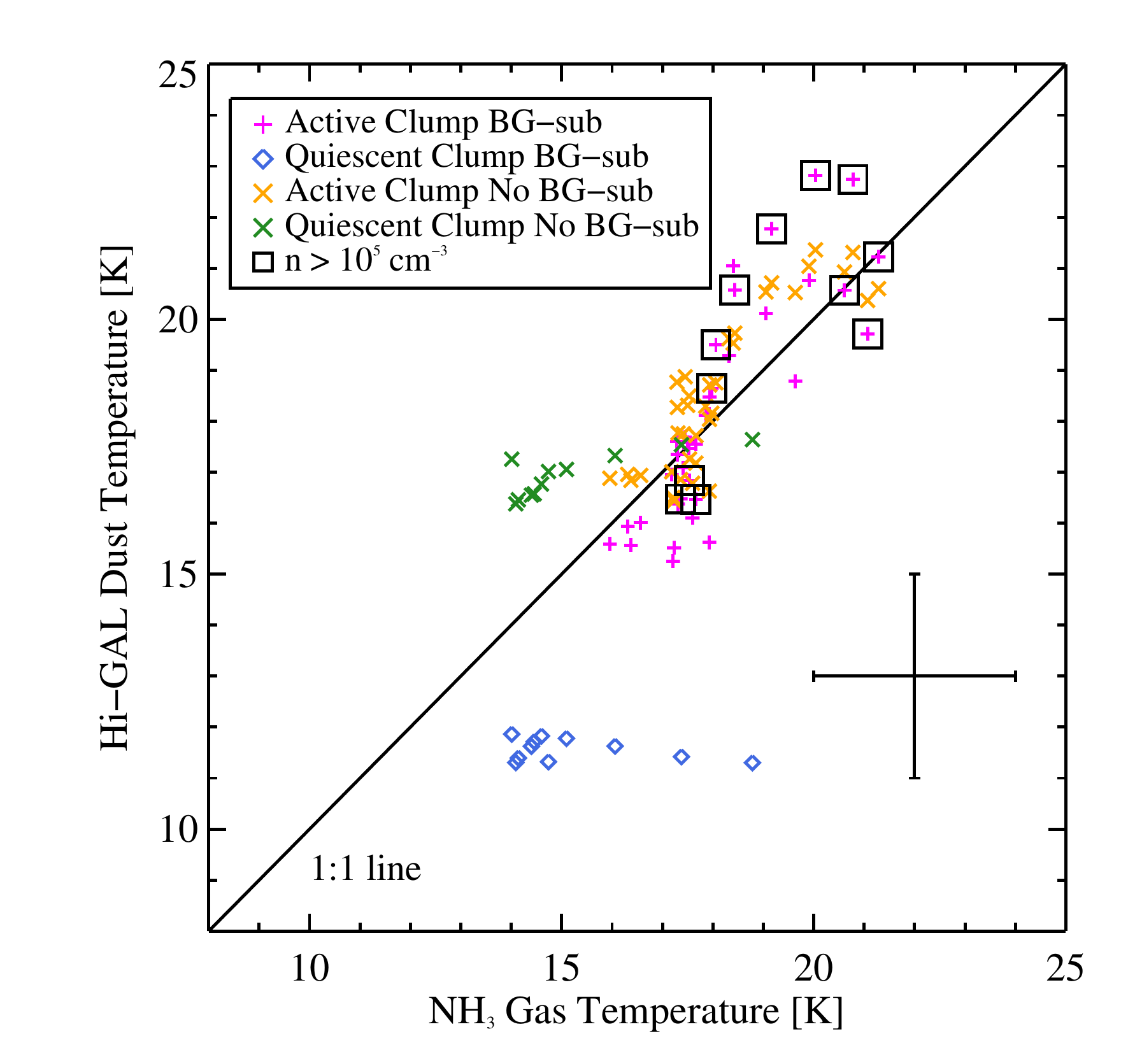}}
\hspace{-0.25in}
\subfigure{
\includegraphics[width=0.49\textwidth]{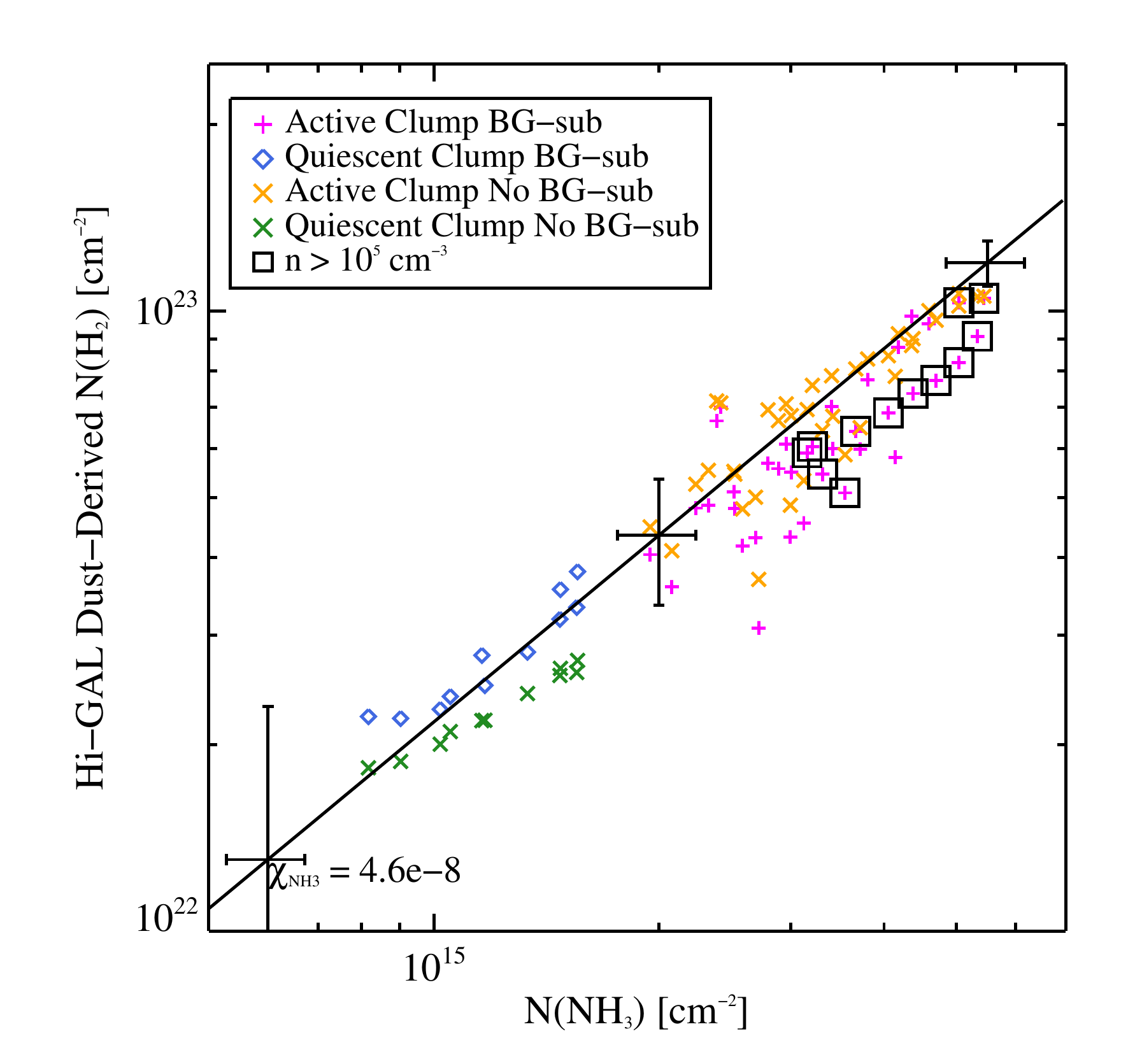}} \\
\caption{A comparison of the Hi-GAL dust-derived temperatures and column densities with those derived from \nh3 on the VLA shows good correlation except for the gas and dust temperatures in the quiescent clump.  The magenta and blue points show the active and quiescent clump pixels, respectively, which have had the background subtracted and have been used throughout the text, while the orange and green points show the active and quiescent clump pixels, respectively, which have not had the background subtracted for comparison.  \textit{Left:} The dust vs. gas temperature for both clumps.  The two show reasonable correlation, though with much scatter, for the active clump and \textit{no} correlation for the quiescent clump.  A typical model fit error bar (2 K in both T$_{dust}$ and T$_{gas}$ is shown in the bottom right).  See the discussion in \S \ref{sec:comp}.
\textit{Right:} The column densities are reasonably well-correlated over a wide range.  A typical model fit error bar (10\% in N(\nh3), 1$\times$10$^{22}$ cm$^{-2}$ in N(H$_{2}$) is shown at three points throughout the plot (to give a sense of errors in the log-log plot).}
\label{fig:higal_nh3comp}
\end{figure*}

%from higal_nh3comppros
\begin{figure*}
\centering			
\includegraphics[width=0.85\textwidth]{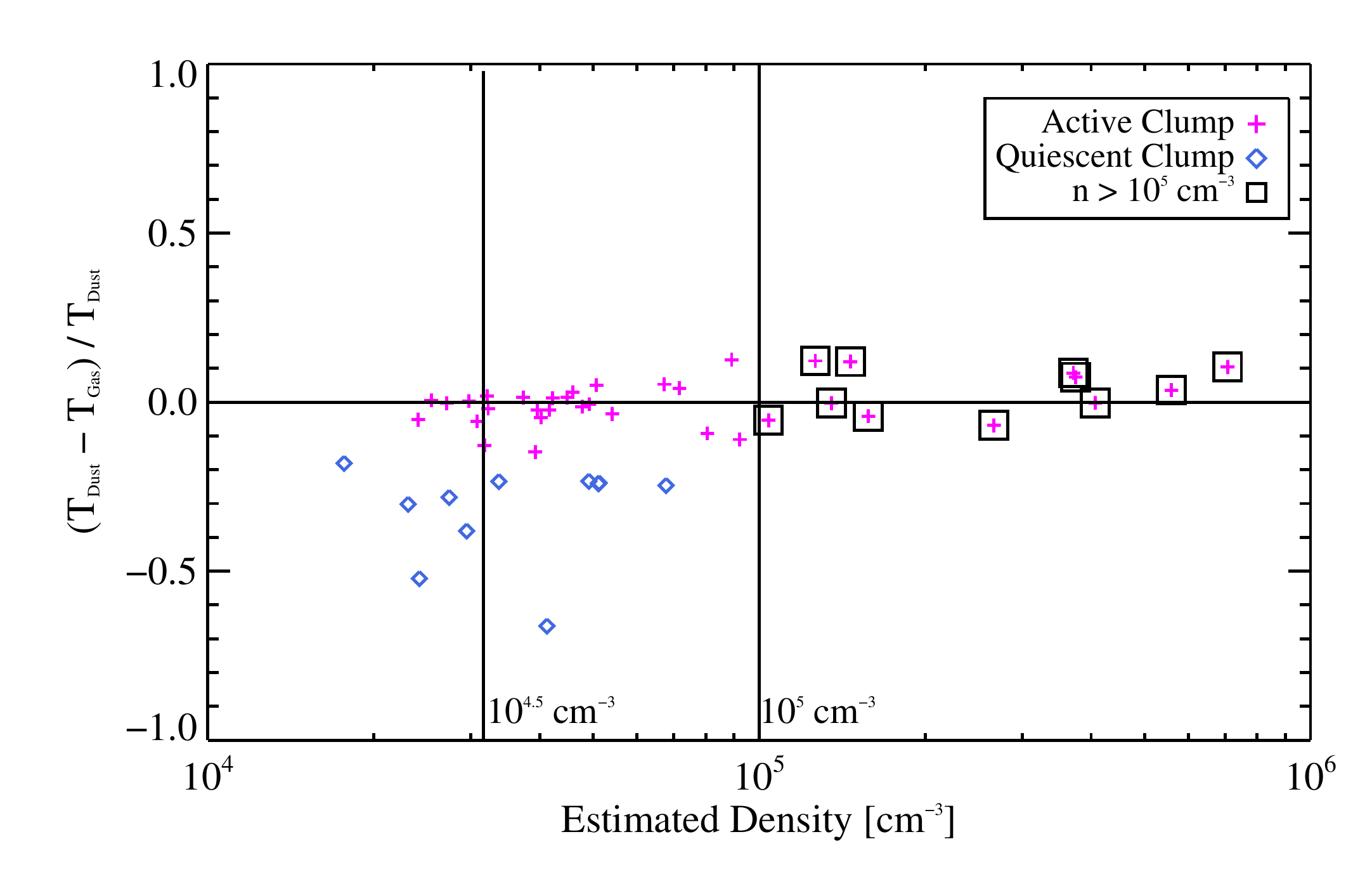}\\
\caption{Estimated volume densities plotted against the fractional difference between the Hi-GAL dust and \nh3 gas temperatures.  The dust and gas are in decent agreement (within about 20\%) in the active clump, however they show disagreement in the quiescent clump, which may indicate a lack of dust/gas coupling or the limits of the dust temperature derivation algorithm in this regime.  This plot agrees well with the cluster-scale simulation of a star-forming region by \citet{mar12} at a time-step of about 0.5 Myr.  The volume densities are estimated as in \S \ref{sec:comp}.  Two vertical lines at n=10$^{4.5}$ and 10$^{5}$ cm$^{-3}$ are plotted to indicate typical densities above which the dust and gas are coupled \citep{gol01}.  }
\label{fig:density_ratio}
\end{figure*}
%also comp_density_ratio_plot-eps-converted-to.pdf

% Depletion Plot
\begin{figure*}
  \centering
     \includegraphics[width=0.75\textwidth]{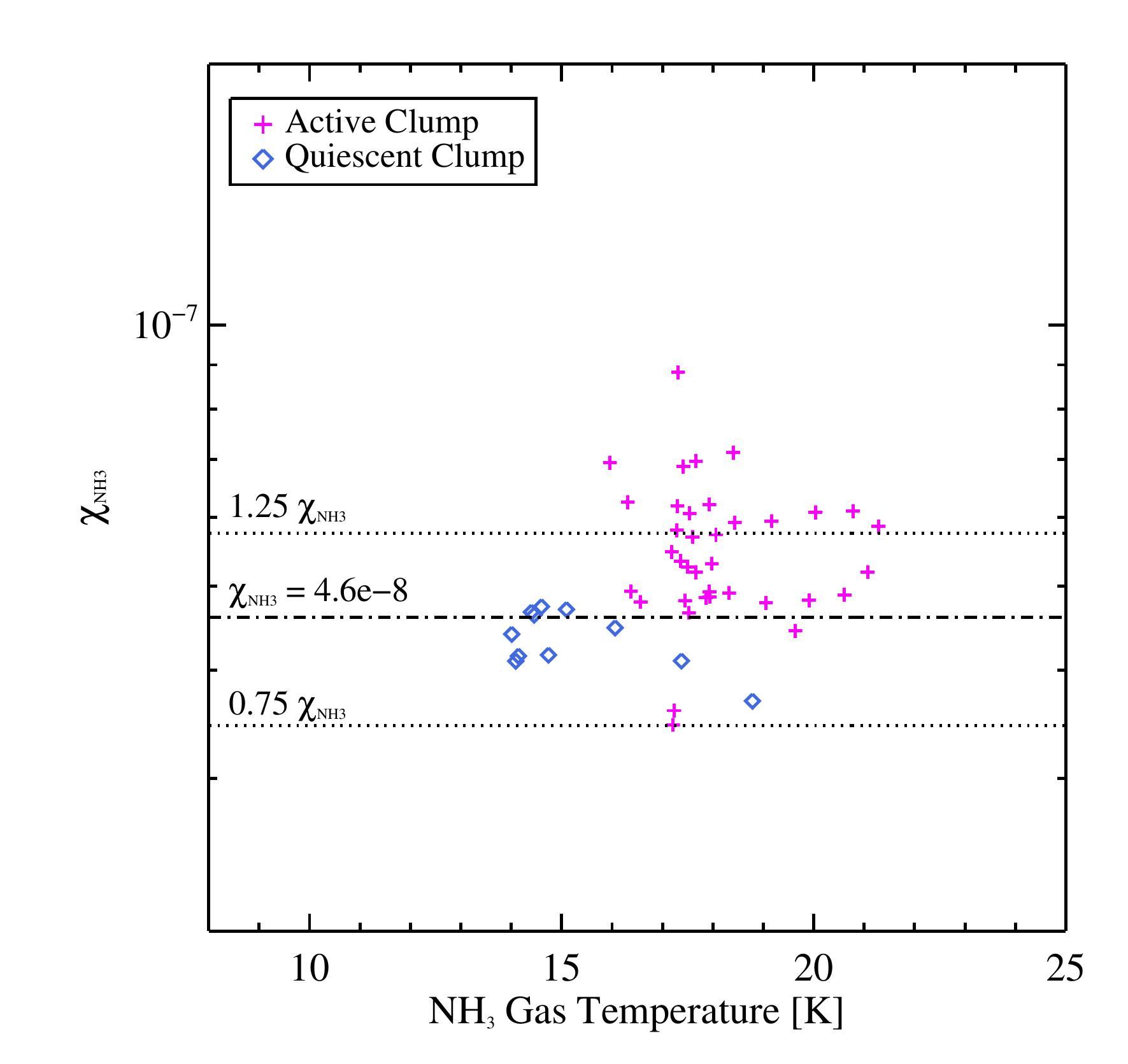}
  \caption{A plot depicting the potential effect of depletion in the quiescent clump, which we estimate to be of order 30\% or less.  This plot shows the ratio of N(\nh3) to Hi-GAL N(H$_{2}$) for each point in the clumps (active in magenta crosses, quiescent in blue diamonds) vs. the associated \nh3 gas temperature.  We adopt an abundance of \abnh3 = 4.6 $\times$ 10$^{-8}$, and we see that most of the active clump points fall at slightly higher abundances and temperatures while the points in the quiescent clump are generally colder and have a slightly lower derived abundance.
 }
  \label{fig:depletion}
\end{figure*}

\section{A Comparison of the Dust and Gas Properties}
\label{sec:comp}
We present a comparison of Hi-GAL dust-derived temperatures and column densities with those derived from \nh3 observations on the VLA in Figures \ref{fig:higal_nh3comp} and \ref{fig:density_ratio} as described in \S \ref{sec:abundance}.
The dust and gas column densities (for the remainder of the text, dust properties refer to those derived with Hi-GAL and gas properties refer to those derived with \nh3 on the VLA, except where stated otherwise) are well correlated over about an order of magnitude.  The derived abundance is slightly higher (about 30\%) in the active than the quiescent clump (see e.g., Figures \ref{fig:higal_nh3comp},  \ref{fig:depletion}),  potentially indicating some small amount of depletion in the quiescent clump.  If the differences in derived abundances from the quiescent clump to the active clump is due to depletion of \nh3 onto dust grains, then the average depletion is of order 30\% or less, see Figure \ref{fig:depletion}.

We estimate volume densities by assuming that the line-of-sight structure sizes are about the same as the observed plane-of-the-sky structure sizes.  We divide both the quiescent and active clump maps into ``core" and ``filament" pixels and use appropriate plane-of-the-sky structure sizes for each to translate the \nh3 column densities into volume densities.  The size estimates used are approximate and reflect the effective diameters from a 2-D Gaussian fit \citep[derived and explained in detail in][]{bat14b}.  The active cores are about 6\arcsec~(0.16 pc) in diameter while the filament surrounding the active cores is about 10\arcsec~(0.27 pc) in diameter.  The quiescent core is about 9\arcsec~(0.24 pc) and its surrounding filament is about 6\arcsec~(0.16 pc) in diameter.  These filaments are wider than the ``universal" filament width found in the Gould Belt by \citet{arz11}.

The estimated volume densities are plotted versus the gas to dust temperature ratio in Figure \ref{fig:density_ratio}, as well as with the boxed symbols in Figure \ref{fig:higal_nh3comp}.  The gas and dust are expected to be coupled above about 10$^{4.5}$ or 10$^{5}$ cm$^{-3}$ \citep{gol01, you04}.  The dust and gas temperatures agree reasonably well (within about 20\%) above n=10$^{5}$ cm$^{-3}$, however the scatter in the temperature ratio does not show any dependence on derived density.  We note that the average densities for the quiescent clump are lower, and typically near or below the threshold density for gas and dust to be coupled.  

While the dust and gas temperature agree within about 20\% in the active clump, the dust and gas temperatures seem completely uncorrelated in the quiescent clump.  We suggest that the dust and gas are not well-coupled in the quiescent clump and that the dust can cool more efficiently than the gas and so is at a uniformly lower temperature, while the gas temperature is higher and variable with location.  Additionally, this indicates that the interplay between gas and dust heating/cooling is not simple in these young star-forming clumps, even at 10$^{4.5}$ cm$^{-3}$.  Alternatively, this offset could be explained if the gas and dust tracers are probing different layers, though we think this is unlikely due to the high critical density of \nh3 and the fact that the interior (as probed more effectively by \nh3) should be cooler in the quiescent clump than the exterior (as probed by dust emission).  It should be noted that the absolute agreement in the quiescent clump is better (they agree within about 20\%) if the Hi-GAL background is not subtracted (see Figure \ref{fig:higal_nh3comp}), however, the dust temperature variation is equally flat with changing gas temperature.  This analysis should be repeated in a region where the Hi-GAL background is less significant, as we may be probing a regime in which the background subtracted modified blackbody fits break down.

The dust and gas temperature vs. density plotted in Figure \ref{fig:density_ratio} shows remarkable similarity to the youngest instance (about 1.6 of the cloud free-fall collapse time or 0.5 Myr) of the simulated cluster-forming region of \citet{mar12} (top left of their Figure 8).  The dust and gas temperatures were calculated individually in this SPH simulation of a cluster-scale ($\sim$ 1 pc, 1000 \Msun) star-forming region.  Just when the gas clump has begun to form a few stars (about 1.6 t$_{ff}$ or 0.5 Myr), the dust is very cold, as it has not yet been significantly heated by the forming stars, while the gas is slightly warmer due to heating from cosmic rays.  The agreement of our observations with this simulation suggests that simulating gas and dust temperatures separately, especially at early times and densities n $<$ 10$^{5}$ cm$^{-3}$ may be important.  This agreement lends some support for a lifetime for the currently observed and previously existing IRDC phase of order 0.5 Myr (similar to the 0.6 - 1.2 Myr starless lifetime found by Battersby et al., in prep.)

\section{Comparison with the Single-Dish GBT Data}	
\label{sec:gbt}
The gas temperatures and column densities found here agree with the single-dish GBT results found by \citet{dun11b} in a survey of \nh3 toward BGPS clumps.  The VLA results presented here have a resolution roughly 10 times better than the BGPS, and are able to detect smaller, higher density regions within the large-scale BGPS clump.  

In the quiescent clump, the VLA observations returned a gas kinetic temperature of 8-15 K for the filament and 13 K for both the Main (Core 1) and Secondary cores (Cores 2, 3, and 4).  The single-dish data in this region centered on the peak of BGPS source number 4901, which is very close to Core 1 in the VLA data.  The single-dish data returned a gas kinetic temperature of 14.9 K, which is slightly warmer than the VLA data because the GBT beam also included a significant amount of the surrounding warmer filament gas within the 31\arcsec~beam.  

In the active clump, the hot core and cold core complexes, as well as the UCHII region all fall within a single BGPS source (BGPS source number 4916 in v1 of the catalog).  In particular, the hot core complex corresponds to the position of the single-dish data.  The single-dish gas temperature is slightly lower than the VLA-derived gas temperature (30.4 K and 35-40 K, respectively).  This difference can be explained in the same way as the difference in the quiescent clump:  the single-dish observations include some of the surrounding cooler gas which lowers the derived kinetic temperature.  Although the exact numbers differ between this study and the single-dish study, the results are consistent.  This comparison highlights the hierarchical nature of star formation, showing increasing complexity and sub-structure down to our resolution limit.

%%%%%%%%%%%%%%
%% BGPS Comp Figures %%
%%%%%%%%%%%%%%
% Made using 'make_bgpscomp_histogram.pro'
%
% Updated Feb. 6, 2013
%
%%%
\begin{figure*}
  \centering
    \includegraphics[width=1\textwidth]{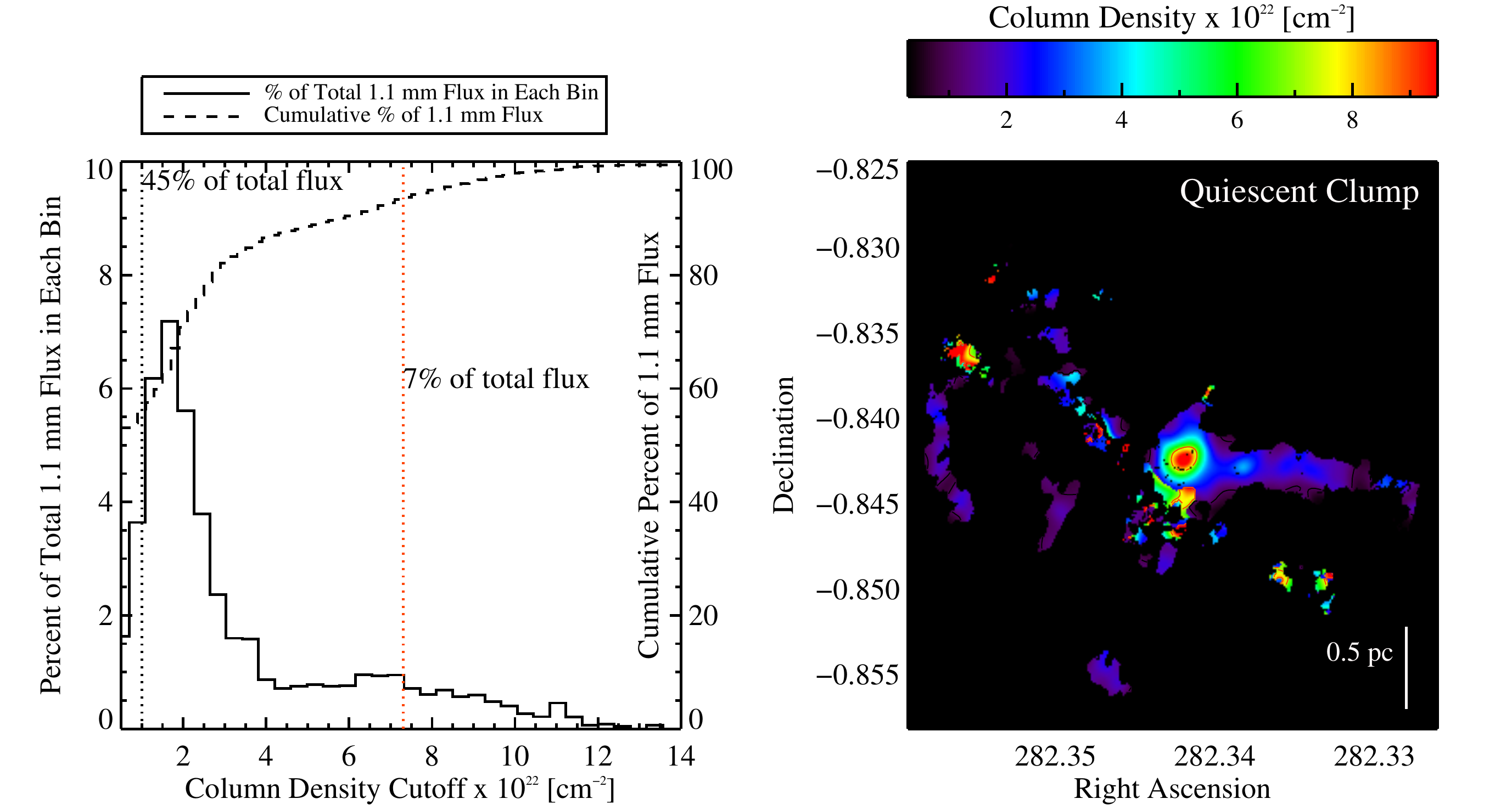}
  \caption{A depiction of the origin of 1.1 mm BGPS flux in the quiescent clump, assuming it has the same temperature and column density structure as the \nh3 on small-scales (see text \S \ref{sec:bgps})   The plot on the left shows the fractional (solid) and cumulative (dashed) distribution of 1.1 mm flux as a function of column density.  A few lines are shown for reference at [1, 7.3, 21] $\times$ 10$^{22}$ cm$^{-2}$ (black, orange, purple, see text \S \ref{sec:bgps}).  The image on the right shows the column density map with contours at the reference column densities in corresponding colors.  In the quiescent clump 55\% of the total 1.1 mm flux is below N(H$_{2}$) = 1 $\times$ 10$^{22}$ cm$^{-2}$ and is filtered out by the interferometer, while 45\% of the flux comes from dense filaments (black dotted line, N(H$_{2}$) $>$ 1 $\times$ 10$^{22}$ cm$^{-2}$), 7\% of the flux arises from massive star forming cores \citep[orange dotted line, N(H$_{2}$) $>$ 7.3 $\times$ 10$^{22}$ cm$^{-2}$, ][]{kau10}, and 0\% of the flux arises from cores above the theoretical threshold for forming massive stars \citep[purple dotted line in Figure \ref{fig:bgpscomp_active} not shown here, N(H$_{2}$) $>$ 21 $\times$ 10$^{22}$ cm$^{-2}$, ][]{kru08}.    
  }
  \label{fig:bgpscomp_quiescent}
\end{figure*}

\begin{figure*}
  \centering
     \includegraphics[width=1\textwidth]{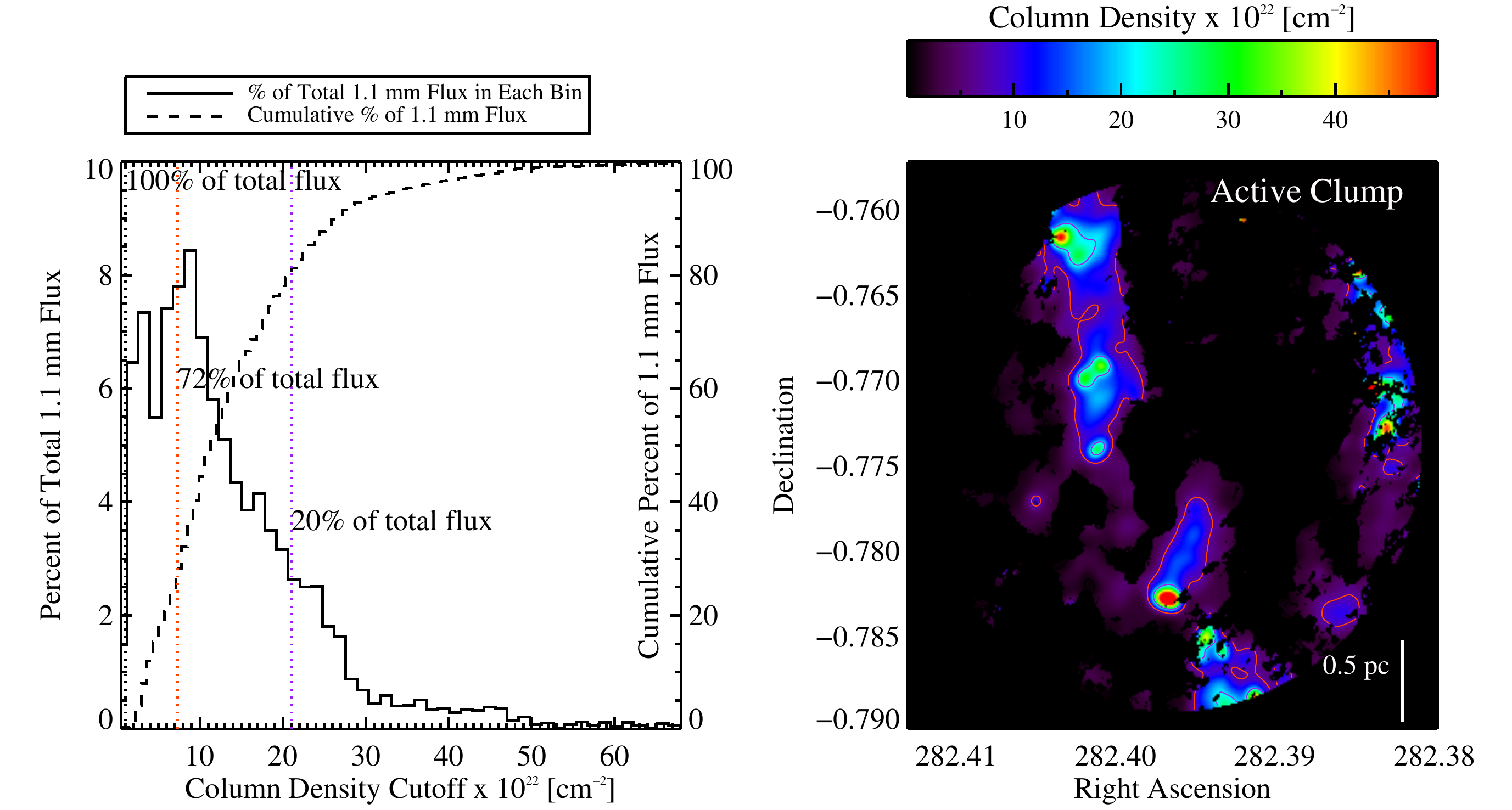}
  \caption{Same as Figure \ref{fig:bgpscomp_quiescent}, but for the active clump.  A depiction of the origin of 1.1 mm BGPS flux in the active clump, assuming it has the same temperature and column density structure as the \nh3 on small-scales (see text \S \ref{sec:bgps}).  In the active clump none of the emission has been filtered out (simulated 1.1 mm BGPS flux is consistent with total real 1.1 mm BGPS flux), meaning that 100\% of the flux originates from dense filaments (black dotted line), 72\% from massive star forming cores \citep[orange dotted line, N(H$_{2}$) $>$ 7.3 $\times$ 10$^{22}$ cm$^{-2}$, ][]{kau10}, and 20\% from the densest massive star-forming cores \citep[purple dotted line, N(H$_{2}$) $>$ 21 $\times$ 10$^{22}$ cm$^{-2}$, ][]{kru08}.
  }
  \label{fig:bgpscomp_active}
\end{figure*}

\section{The High-Resolution Origin of Sub-mm Clump Emission}
\label{sec:bgps}
The high-resolution \nh3 data from the VLA allows the unique opportunity to explore the properties of dense gas on sub-pc scales.  The Bolocam Galactic Plane Survey (BGPS) on the other hand allows for an investigation of these properties across the Galaxy, but on $\gtrsim$ 1 pc scales (33\arcsec~resolution corresponds to clumps on the near side of the Galaxy or whole clouds on the far side).  A comparison of this high resolution \nh3 data with large-scale BGPS data allows an opportunity to address the small-scale origin of the millimeter flux throughout the Galaxy. 

In this particular IRDC, the contribution of free-free emission to the millimeter flux is negligible \citep{bat14b}, as the only radio continuum source (G$32.03+0.05$) observed at 1, 6 and 20 cm is not coincident with the millimeter emission; the source is evolved enough to have blown out most of the dense gas and dust.  In order to determine the relative contributions of the cores and filaments to the millimeter flux, we invert Eq \ref{eq:dust_col} in \S \ref{sec:dust_col},  the equation to derive the dust column density from millimeter flux, and solve it for the millimeter flux using the column density and temperature maps derived from the \nh3 data.  This gives us a simulated map of 1.1 mm flux on sub-pc scales.  This, of course, assumes a tight coupling between the dense gas and dust.  If the gas and dust are less well-coupled, especially in the lower density filaments, then we are overestimating the flux from the filament (if the dust is colder than we are assuming), and a higher fraction of the millimeter emission arises from the dense gas cores.  In the quiescent clump, where we think that the true dust temperature is lower than the measured gas temperature (12 K vs. 15 K), the simulated fluxes at 1.1mm would be about 70\% of what is reported here.  The fractions, however, remain unchanged as they are relative to the total.

In Figures \ref{fig:bgpscomp_quiescent} and \ref{fig:bgpscomp_active} we measure which structures are responsible for the observed pc-scale sub-mm emission observed at 1.1 mm with the BGPS.  We compare the simulated (forward-modeled) 1.1 mm flux with the real 1.1 mm BGPS flux and find that in the quiescent clump, about 55\% of the global real BGPS 1.1 mm flux is filtered out by the interferometer / masking, meaning that only about 45\% of the real BGPS 1.1 mm flux arises from dense filaments with N(H$_{2}$) $>$ 10$^{22}$ cm$^{-2}$.  In the active clump, the gas is more compact and all of it appears to arise from structures with N(H$_{2}$) $>$ 10$^{22}$ cm$^{-2}$ (i.e. the simulated sub-pc 1.1 mm flux matches the real pc-scale 1.1 mm flux).  About 7\% vs. 72\% of the total 1.1 mm flux arises from massive star-forming dense cores for the quiescent and active clumps respectively, where massive star-forming cores refers to being above the massive star-forming threshold from \citet{kau10} which corresponds very roughly to N(H$_{2}$) $>$ 7.3 $\times$ 10$^{22}$ cm$^{-2}$ for one resolution element in our maps.  About 0\% vs. 20\% of the total 1.1 mm flux arises from the densest cores with $\Sigma$ $>$ 1 g cm$^{-2}$ in the quiescent and active clumps respectively, the theoretical threshold for the formation of massive stars from \citet{kru08}, corresponding to about N(H$_{2}$) $>$ 21 $\times$ 10$^{22}$ cm$^{-2}$.

In summary, sub-mm BGPS clumps show a variety of sub-pc structure and the origin of 1.1 mm emission can be from diffuse clumps, dense filaments, and massive star-forming cores.  In some clumps (like the quiescent clump), over 50\% of the flux arises from the diffuse clump, about 40\% from the dense filament, and only about 10\% from massive star-forming cores.  In other clumps (like the active clump), all of the sub-mm flux arises from dense filamentary structures with N(H$_{2}$) $>$ 10$^{22}$ cm$^{-2}$ and about 75\% from massive star-forming cores, in a range of evolutionary states.  Sub-mm dust clumps probe a range of density structures and star-forming evolutionary states.  The fact that the more evolved clump contains a significantly larger dense gas fraction than the quiescent clump is an interesting result.  This fraction may be a meaningful measure of the global collapse of a clump, however, a larger sample is required to address this correlation.

\section{Conclusions}
\label{sec:conc}
We explore the relationship between dust and gas derived physical properties in a massive star-forming IRDC, G$32.03+0.05$ showing a range of evolutionary states.  The gas properties (temperature and column density) were derived using radiative transfer modeling of three inversion transitions of para-\nh3 on the VLA, while the dust properties (temperature and column density) were derived with cirrus-subtracted modified blackbody fits to Herschel data.  We derive an \nh3 abundance and compare different tracers of column density (dust emission and extinction and gas emission).  The gas and dust temperatures agree well in the active clump, but the disagreement in the quiescent clump calls into question either the reliability of dust temperatures in this regime or the assumption of tight gas and dust coupling.  We also explore the high-resolution ($\sim$ 0.1 pc) origin of pc-scale dust emission by forward modeling the gas temperature and column densities to derive 1.1 mm fluxes and comparing with the observed BGPS 1.1 mm fluxes.

 \begin{itemize}
\item \textit{\nh3~Abundance:}  A comparison of the \nh3 column density with those derived from various independent tracers exemplifies some of the uncertainties involved in such measurements, as the systematic variations between these tracers is about a factor of two.  The Hi-GAL dust continuum data show a good correlation with the \nh3 from which we derive an abundance of \abnh3 = 4.6 $\times$ 10$^{-8}$, in agreement with previous single-dish observations.  This agreement indicates that both the gas and dust are clumpy on scales smaller than about 60\arcsec~(the largest angular size to which our \nh3 observations are sensitive).

\item \textit{Gas and Dust Coupling:}  The gas and dust column densities show good agreement.  Depletion of \nh3 in the quiescent clump is of order 30\% or less.  The dust and gas temperatures are scattered, but agree reasonably well for the active clump (within about 20\%).  The quiescent clump, however, shows no correlation between dust and gas temperature.  At lower densities in the quiescent clump, the two may not be well-coupled, or we may be probing a regime in which the dust or \nh3 temperature estimates break down.  A comparison of these temperatures and densities agrees well with a cluster-scale star formation
simulation~\citep{mar12} just before the stars begin to turn on (about 0.5 Myr).  The agreement of our observations with this simulation suggests that simulating the gas and dust heating and cooling processes individually in young star-forming regions may be important.

\item \textit{The Origin of Dust Continuum Emission:}  Forward modeling of the \nh3 data (temperatures and column densities) to produce millimeter fluxes and comparing these with BGPS millimeter fluxes reveals that millimeter dust continuum observations, such as the BGPS, Hi-GAL, and ATLASGAL, probe hot cores, cold cores, as well as the dense filaments from which they form.  The millimeter flux is not dominated by a single hot core, but rather, is representative of the cold, dense gas as well.
We estimate that the quiescent clump is dominated by diffuse emission, much of which (about 55\%) is filtered out by the interferometer in our VLA measurements.  In this clump, the cold, dense filament comprises about 45\% of the total flux.
Only about 7\% of the total 1.1 mm flux arises from massive star-forming cores in the quiescent clump.  The active clump is dominated by high density gas, both hot and cold, in the form of dense filaments (100\% of the flux from material with N(H$_{2}$) $>$ 10$^{22}$ cm$^{-2}$) and massive star-forming cores (72\% of the flux from massive star-forming cores; N(H$_{2}$) $>$ 7.3 $\times$ 10$^{22}$ cm$^{-2}$ and 20\% from material with N(H$_{2}$) $>$ 2.1 $\times$ 10$^{23}$ cm$^{-2}$).

\end{itemize}

\acknowledgments

We thank the referee for his or her helpful comments which have helped to improve the quality of this manuscript.  We thank Hugo Martel and Neal Evans for assistance comparing our results with their cluster-scale simulations.  
This work has made use of the GLIMPSE and MIPSGAL surveys, and we thank those teams for their help and support.  We would like to thank the staff at VLA for their assistance.  The National Radio Astronomy Observatory is a facility of the National Science Foundation operated under cooperative agreement by Associated Universities, Inc.  This publication makes use of molecular line data from the Boston University-FCRAO Galactic Ring Survey (GRS). The GRS is a joint project of Boston University and Five College Radio Astronomy Observatory, funded by the National Science Foundation under grants AST-9800334, AST-0098562, AST-0100793, AST-0228993, \& AST-0507657.  This work has made use of ds9 and the Goddard Space Flight Center's IDL Astronomy Library.  Data processing and map production of the Herschel data has been possible thanks to generous support from the Italian Space Agency via contract I/038/080/0. Data presented in this paper were also analyzed using The Herschel interactive processing environment (HIPE), a joint development by the Herschel Science Ground Segment Consortium, consisting of ESA, the NASA Herschel Science Center, and the HIFI, PACS, and SPIRE consortia.

\bibliography{/Users/battersby/Dropbox/references1}{}

\end{document}